\documentclass[12pt]{article}
\usepackage{t1enc}
\usepackage[utf8]{inputenc}
\usepackage[english]{babel}
\usepackage{graphicx}
\usepackage{amssymb}
\usepackage{amsmath}
\usepackage{amsthm}
\usepackage{braket}
\usepackage{mathtools}
\usepackage{bm}
\usepackage{bbm}
\usepackage{caption}

\def\UN{\mathbf{1}}

\def\B{{\cal B}}
\def\CO{{\mathbb C}}
\def\RE{{\mathbb R}}

\def\qed{\ \vrule height 5pt width 5pt depth 0pt}
\def\cros{\raise1.9pt\hbox{$\scriptscriptstyle
          >$}\!\raise1.5pt\hbox{$\scriptstyle\triangleleft\,$}}

\def\C{{\cal C}}

\def\s{\sigma}
\def\P{{\cal P}}

\def\B{{\cal B}}
\def\H{{\cal H}}

\def\w{\wedge}

\theoremstyle{definition}
\theoremstyle{definition}
\theoremstyle{definition}
\theoremstyle{definition}\newtheorem{Lemma}{Lemma}
\newcommand{\noi}{\vspace{0.1in} \noindent}
\newcommand{\noii}{\vspace{0.2in} \noindent}

\title{\bf Quantum mechanics as a noncommutative representation of classical conditional probabilities}
\author{\textit{Gábor Hofer-Szabó}\thanks{Research Center for the Humanities, Budapest, Hungary; email: szabo.gabor@btk.mta.hu}}
\date{}

\begin{document}
\maketitle

\begin{abstract}
The aim of this paper is to analyze the reconstructability of quantum mechanics from classical conditional probabilities representing measurement outcomes conditioned on measurement choices. We will investigate how the quantum mechanical representation of classical conditional probabilities is situated within the broader frame of noncommutative representations. To this goal, we adopt some parts of the quantum formalism and ask whether empirical data can constrain the rest of the representation to conform to quantum mechanics. We will show that as the set of empirical data grows conventional elements in the representation gradually shrink and the noncommutative representations narrow down to the unique quantum mechanical representation.
\vspace{0.1in}

\end{abstract}

\section{Introduction}

In the quantum information theoretical paradigm one is usually looking for the reconstruction of quantum mechanics from information-theoretic first principles (Hardy, 2008; Chiribella, D'Ariano and Perinotti, 2017). This approach has produced many fascinating mathematical results and greatly contributed to a better understanding of the complex formal structure of quantum mechanics. As a top-down approach, however, its prime aim was to clarify the relation of the theory to higher-order (rationality, information-theoretic, etc.) principles and payed less attention to the ``legs'' of the theory connecting it to experience.

In this paper we take an opposite, bottom-up route and ask---in the spirit of the good old empiricist tradition---as to how the theory can be reconstructed not from first principles but from experience. More precisely, we will ask whether we can reconstruct the formalism of quantum mechanics from using simply classical conditional probabilities.

Why classical conditional probabilities?

Quantum mechanics as a probabilistic theory provides us quantum probabilities for certain observables. The question is how to connect these quantum probabilities to experience. The correct answer is that the probabilities provided by the Born rule should be interpreted as \textit{classical conditional probabilities}. They are \textit{classical} since they are nothing but the long-run relative frequency of certain measurement outcomes explicitly testable in the lab; and they are \textit{conditional} on the fact that a certain measurement had been chosen and performed (E. Szabó, 2008). For example, the quantum probability of the outcome ``spin-up'' in direction $z$ is the relative frequency of the outcomes ``up''---but not in the statistical ensemble of \textit{all} measurement outcomes (which may also comprise spin measurements in \textit{other} directions) but only in the \textit{sub}ensemble when spin was measured in direction $z$. 

What does it mean to reconstruct quantum mechanics from classical conditional probabilities?

First note that all we are empirically given are classical conditional probabilities. The question is how to represent these empirical data. As it was hypothesized in (E. Szabó 1995) and shown---under some specific conditions---in (Bana and Durt 1997), (E. Szabó 2001) and (Rédei 2010) classical conditional probabilities conforming to the probabilistic predictions of quantum mechanics need \textit{not} necessarily be represented in the formalism of quantum mechanics. The so-called ``Kolmogorovian Censorship Hypothesis'' (or better, Proposition) states that there is always a Kolmogorovian representation of the quantum probabilities \textit{if} the measurement conditions also make part of the representation. Thus, a stubborn classicist will always find a way to represent the empirical content of quantum mechanics in a purely classical framework. 

On the other hand, quantum mechanics has proved to be an extremely elegant and economic representation of these empirical data. It provides a principled representation of an enormous collection of conditional probabilities together with their dynamical evolution. 

Our paper is a kind of interpolation between the two sides. Our strategy will be to accept \textit{some} parts of the quantum mechanical representation of classical conditional probabilities and ask whether the \textit{rest} follows. More precisely, we accept the noncommutative probability theory (see Gudder 1988; Rédei and Summers, 2007) which in our case will boil down to representing observables and states by linear operators. We also adopt the Born rule connecting the quantum probabilities to real-world classical conditional probabilities; and the quantum mechanical representation of measurement settings and measurement outcomes. The only ``free variable'' will be the representation of the \textit{state} of the system. Our main question will then be as to what empirical data ensure that the state of a system is represented by a density operator.

By this strategy we are going to analyze how quantum mechanics is situated within a noncommutative probability theory and to study whether the specific quantum mechanical representation of classical conditional probabilities within this broader frame can be traced back to purely empirical facts or is partly of conventional nature.

In the paper we will proceed as follows. In Section 2 we introduce the general scheme of a noncommutative representation of classical conditional probabilities. In the subsequent three sections we gradually enhance the set of empirical data that is the set of classical conditional probability of measurement outcomes. We ask whether by increasing the set of empirical data the noncommutative representation of these data necessarily narrows down to the quantum mechanical representation or some extra conventional elements are also needed. The empirical situation we are going to represent will be three yes-no measurements in Section 3, $k$ measurements each with $n$ outcomes in Section 4, and finally a continuum set of measurements with $n$ outcomes in Section 5. We will see how the conventional part gradually shrinks as experience grows until the representation finally zooms in on the quantum mechanical representation. We discuss our results in Section 6.

\section{Quantum mechanical and noncommutative representation}

Suppose there is a physical system in state $s$ and we perform a set $\{a_i\}$ $(i\in I)$ of measurements on the system. Denote the outcomes of measurement $a_i$ by $\{A^j_i\}$ $(j\in J)$. Suppose that by repeating the measurements many times we obtain a probability $p_s(A^j_i|a_i)$ that is a stable long-run relative frequency for each outcome $A^j_i$ \textit{given} measurement $a_i$ is performed. Now, quantum mechanics represents these conditional probabilities as it is summarized in the following table:
\begin{table}[h]
\begin{tabular}{ccc} \hline \hline 
Operator assignment: && Born rule: \\ \hline \hline 
\begin{tabular}{rll} 
System & $\longrightarrow$ & $\H$: Hilbert space \\
Measurements: $a_i$ & $\longrightarrow$ & $O_i$: self-adjoint operators \\
Outcomes: $A^j_i$ & $\longrightarrow$ & $P^j_i$: spectral projections of $O_i$ \\ 
States: $s$ & $\longrightarrow$ & $W_s$: density operators \\ 
\end{tabular} 
&&
\begin{tabular}{l} 
$p_s(A^j_i|a_i) = \mbox{Tr}(W_sP^j_i)$ 
\end{tabular} \\ \hline \hline 
\end{tabular}
\caption{Quantum mechanical representation}
\end{table}

\noindent
In the table the different concepts are presented. On the left hand side of the arrow/equation sign stand the empirical concepts to be represented; on the right hand side stand the mathematical representation of the empirical concepts. The two are not to be mixed. Although we do not use ``hat'' to denote operators, throughout the paper we carefully distinguish empirical terms (measurements, outcomes, states) from their representation (self-adjoint operators, projections, density operators). Thus, the physical system under investigation is associated to a Hilbert space $\H$; each measurement $a_i$ is represented by a self-adjoint operator $O_i$; the outcomes $A^j_i$ of $a_i$ are represented by the orthogonal spectral projections of $O_i$; and the state $s$ of the system is represented by a density operator $W_s$, a self-adjoint, positive semidefinite operator with trace equal to 1. In the second column the mathematical representation is connected to experience by the Born rule: the representation is correct only if the quantum mechanical trace formula $\mbox{Tr}(W_sP^j_i)$ correctly yields the empirical conditional probability $p_s(A^j_i|a_i)$ for any outcome $A^j_i$ of measurement $a_i$ and any state $s$. 

Note the following two facts. First, the trace formula is associated to a \textit{conditional} probability, not to a probability \textit{simpliciter}. This means, among others, that in joint measurements one always needs to combine different measurement conditions. Second, the trace formula is ``holistic'' in the sense that the empirically testable conditional probabilities are associated to the trace of the \textit{product} of two operators, one representing the state and the other representing the measurement. This leaves us with a lot of freedom to account for the same empirical content in terms of operators.

The main question of our paper is whether the above quantum mechanical representation of classical conditional probabilities is constrained upon us if the set of empirical data is large enough or whether we need some extra theoretical, aesthetic etc.\ considerations to arrive at it. In order to decide on this question, we consider first a wider class of representations which we will call \textit{noncommutative representations}. We will then ask whether a noncommutative representation of a set of large enough data is necessarily a quantum mechanical representation. 

What is a noncommutative representation?

Generally, a noncommutative representation is simply an association of measurements and states to linear operators acting on a Hilbert space such that some functional of the representants provides the correct empirical conditional probabilities. Obviously this association can be done in many different ways. In our paper we pick a special noncommutative representation which is very close to the quantum mechanical representation: We retain all the assignments (denoted by $\longrightarrow$) of the above table except the last one. That is we will represent the system by a Hilbert space, the measurements by self-adjoint operators, and the outcomes by the orthogonal spectral projections. We also retain the Born rule connecting the formalism to experience. The only part of the representation which we let vary will be the association of the state of the system to linear operators. That is we do not demand that states should necessarily be represented by density operators. We summarize this scheme in the following table:
\begin{table}[h]
\begin{tabular}{ccc} \hline \hline 
Operator assignment: && Born rule: \\ \hline \hline 
\begin{tabular}{rll} 
System & $\longrightarrow$ & $\H$: Hilbert space \\
Measurements: $a_i$ & $\longrightarrow$ & $O_i$: self-adjoint operators \\
Outcomes: $A^j_i$ & $\longrightarrow$ & $P^j_i$: spectral projections of $O_i$ \\ 
States: $s$ & $\longrightarrow$ & $W_s$: linear operators \\ 
\end{tabular} 
&&
\begin{tabular}{l} 
$p_s(A^j_i|a_i) = \mbox{Tr}(W_sP^j_i)$ 
\end{tabular} \\ \hline \hline 
\end{tabular}
\caption{Noncommutative representation}
\end{table}

\noindent
Obviously, our noncommutative representation is only one special choice among many. One could well take different routes. For example one could demand that the state should be represented by density operators but abandon that the projections representing the outcomes should be orthogonal. Or one could replace the Born rule by another expression connecting the formalism to the world. As said above, the connection of the formalism of quantum mechanics and experience is of holistic nature; one can fix one part of the formalism and see how the rest may vary such that the resulting probabilities are in tune with experience. With respect to our aim which is to see how we are compelled to adopt the quantum mechanical representation by increasing the number of conditional probabilities to be represented, our above choice is just as good as any other.

What we will test in the subsequent sections is whether our noncommutative representation is necessarily a quantum mechanical representation. In other words, we will test whether for any choice of operators representing a certain set of measurements and the outcomes such that the Born rule yields the correct conditional probabilities, the state will necessarily be represented by a density operator. In Section 3 we start off as a warm-up with three measurements; in Section 4 we continue with $k$ measurements; and in Section 5 we end up by uncountably many measurements. It will turn out that the gap between noncommutative and quantum mechanical representation gradually shrinks as the set of empirical data grows.

\section{Case 1: Three yes-no measurements}

Consider a box filled with balls. Denote the preparation of the box by $s$. Suppose you can perform three different measurements on the system; you can measure the \textit{color}, the \textit{size} or the \textit{shape} of the balls. Denote the three measurements as follows:

\noi
\begin{tabular}{rl} 
$a$: & Color measurement \\
$b$: & Size measurement \\
$c$: & Shape measurement 
\end{tabular}

\noi
Suppose that each measurement can have only two outcomes:

\noi
\begin{tabular}{rlcrl} 
$A^+$: & Black && $A^-$: & White  \\ 
$B^+$: & Large && $B^-$: & Small \\
$C^+$: & Round && $C^-$: & Oval
\end{tabular}

\noi
Suppose you pick a measurement, perform it many times (putting the balls always back into the box), and count the probability, that is the long-run relative frequency, of the outcomes. What you obtain is the \textit{conditional} probability of the outcomes \textit{given} the measurement you picked is performed on the system prepared in state $s$:
\begin{eqnarray}
p^\pm_a &:=& p_s(A^\pm|a) \label{p(a)} \\
p^\pm_b &:=& p_s(B^\pm|b) \label{p(b)}  \\
p^\pm_c &:=& p_s(C^\pm|c) \label{p(c)}
\end{eqnarray}
Now, suppose you are going to represent the above empirical facts not in the standard classical probability theory but in a quantum fashion. Since our model contains only two-valued (yes-no) measurements, it suffices to use only a minor fragment of quantum mechanics. Again, we summarize it in a table:
\begin{table}[h]
\begin{tabular}{ccc} \hline \hline 
Operator assignment: && Born rule: \\ \hline \hline 
\begin{tabular}{rll} 
System: & $\longrightarrow$ & $\CO_2$ \\
Color: $a$ & $\longrightarrow$ & $O_a = {\bf a}\bm{\s}$\\
Size: $b$ & $\longrightarrow$ & $O_b = {\bf b}\bm{\s}$\\
Shape: $c$ & $\longrightarrow$ & $O_c = {\bf c}\bm{\s}$\\
Black/White: $A^\pm$ & $\longrightarrow$ & $P^\pm_a = \frac{1}{2}(\UN \pm {\bf a}\bm{\s})$ \\ 
Large/Small: $B^\pm$ & $\longrightarrow$ & $P^\pm_b = \frac{1}{2}(\UN \pm {\bf b}\bm{\s})$ \\ 
Round/Oval: $C^\pm$ & $\longrightarrow$ & $P^\pm_c = \frac{1}{2}(\UN \pm {\bf c}\bm{\s})$ \\ 
State: $s$ & $\longrightarrow$ & $W_s = \frac{1}{2}(\UN + {\bf s}\bm{\s})$ \\
\end{tabular}
&&
\begin{tabular}{c} 
$p^\pm_a = \mbox{Tr}(W_sP^\pm_a) = \frac{1}{2}(1 \pm {\bf sa})$ \\
$p^\pm_b = \mbox{Tr}(W_sP^\pm_b) = \frac{1}{2}(1 \pm {\bf sb})$ \\
$p^\pm_c = \mbox{Tr}(W_sP^\pm_c) = \frac{1}{2}(1 \pm {\bf sc})$
\end{tabular} \\ \hline \hline 
\end{tabular}
\caption{Quantum mechanical representation of three yes-no measurements in $\CO_2$}
\end{table}

\noindent
Here, the Hilbert space associated to the system is the two-dimensional complex space $\CO_2$; and the operators associated to the measurements, outcomes and the state are all self-adjoint operators acting on $\CO_2$. According to this representation, called the Bloch sphere representation, a self-adjoint operator $O_a$ associated to measurement $a$ can be represented by the inner product of a unit vector ${\bf a}= (a_x,a_y,a_z)$ in $\RE^3$ and the Pauli vector $\bm{\s}= (\s_x,\s_y, \s_z)$. The two outcomes $A^\pm$ of measurement $a$ are associated to the spectral projections $P^\pm_a = \frac{1}{2}(\UN \pm {\bf a}\bm{\s})$ of $O_a$, where $\UN$ is the two-dimensional identity operator. Finally, the density operator $W_s$ associated to the state $s$ of the system is of the form $W = \frac{1}{2}(\UN + {\bf s}\bm{\s})$, where ${\bf s}= (s_x,s_y,s_z)$ is in the unit ball $B=\{{\bf r} \in \RE^3: |{\bf r}| \leqslant 1\}$ of $\RE^3$. If $|{\bf s}| = 1$, then $s$ is said to be a  pure state, otherwise a mixed state. Again, the empirical content of the representation is ensured by the Born rule which in this two-dimensional case boils down to the inner product: $p^\pm_a = \frac{1}{2}(1 \pm {\bf sa})$. (Similarly for $b$ and $c$.)

Now, to give a quantum mechanical representation for the above situation we need to associate the three measurements to three Bloch vectors and the state of the system to a fourth Bloch vectors (either unit or smaller) such that the Born rule (the trace formula) yields the pre-given conditional probabilities (\ref{p(a)})-(\ref{p(c)}). Thus, assign to each measurement a unit vector in $\RE^3$:
\begin{eqnarray}
\{a,b,c\} \mapsto \{{\bf a}, {\bf b}, {\bf c}\} \label{assign}
\end{eqnarray}
Suppose that the vectors ${\bf a}$, ${\bf b}$ and ${\bf c}$ are linearly independent. First, we show that given three pairs of empirical conditional probabilities $p^\pm_a$, $p^\pm_b$ and $p^\pm_c$ and also the assignment (\ref{assign}), the operator $W_s$ associated to the state $s$ gets uniquely fixed. Schematically,
\begin{eqnarray*}
p^\pm_a, \, p^\pm_b, \, p^\pm_c \quad \& \quad {\bf a}, \, {\bf b}, \, {\bf c} \quad \Longrightarrow \quad W_s
\end{eqnarray*}
To see this, observe that any linear operator acting on $\CO_2$ can be written as
\begin{eqnarray*}
W_s= s_0 \UN + {\bf s}\bm{\s} +i(s'_0 \UN + {\bf s'}\bm{\s})
\end{eqnarray*}
where $s_0, s'_0 \in \RE$ and ${\bf s}, {\bf s'} \in \RE^3$. Now, applying the Born rule to the three measurements we get:
\begin{eqnarray*}
&&p^\pm_a = \mbox{Tr}(W_sP^\pm_a) = s_0 \pm {\bf sa} + i(s'_0 \pm {\bf s'a}) \\
&&p^\pm_b = \mbox{Tr}(W_sP^\pm_b) = s_0 \pm {\bf sb} + i(s'_0 \pm {\bf s'b}) \\
&&p^\pm_c = \mbox{Tr}(W_sP^\pm_c) = s_0 \pm {\bf sc} + i(s'_0 \pm {\bf s'c}) 
\end{eqnarray*}
which, assuming that $p^\pm_a$, $p^\pm_b$ and $p^\pm_c$ are real and ${\bf a}$, ${\bf b}$ and ${\bf c}$ are linearly independent, yield 
\begin{eqnarray*}
s_0 = \frac{1}{2} \quad \quad s'_0 = 0 \quad \quad {\bf s}'={\bf 0} 
\end{eqnarray*}
and hence
\begin{eqnarray*}
&&p^\pm_a = \frac{1}{2} \pm {\bf sa} \\
&&p^\pm_b = \frac{1}{2} \pm {\bf sb}  \\
&&p^\pm_c = \frac{1}{2} \pm {\bf sc} 
\end{eqnarray*}
the solution of which is $W_s = \frac{1}{2}(\UN + {\bf s}\bm{\s})$ with
\begin{eqnarray*}\label{sol}
{\bf s} = \frac{(p^+_a -\frac{1}{2})({\bf b \times c}) + (p^+_b -\frac{1}{2})({\bf c \times a}) + (p^+_c -\frac{1}{2})({\bf a \times b})}{{\bf a}\cdot ({\bf b \times c})}
\end{eqnarray*}
where $\times$ is the cross product. (The linear independence of ${\bf a}$, ${\bf b}$ and ${\bf c}$ is needed for the triple product in the denominator not to be zero.)

\noi
This is a well-known result. Since the late 60s and early 70s there has begun an intensive research for the \textit{empirical} determination of the state of a quantum system. In a series of papers Band and Park (1970, 1971) have extensively investigated how the expectation value of certain observables determine the state of a system. They investigated the minimal number of observables, called the \textit{quorum}, needed for such state determinations; the structure and geometry of this set; and many other important features. The study of the \textit{quorum} has become an eminent research project also in the new quantum informational paradigm. Quantum tomography, quantum state reconstruction, quantum state estimation etc.\ all follow the same path: they start from a set of observables and aim to end up with a more-or-less fixed state using empirical input (see for example (D’Ariano, Maccone and Paris, 2001)).

However, all these endeavors have a common pre-assumption, namely that \textit{the association of measurements to operators is already settled}. They all start from a set of operators and (by means of a set of empirical probabilities) aim to reconstruct the quantum state of a system. But an operator is not a measurement but only a representation of a measurement. Calling operators \textit{observables} overshadows the fact that the operators are already on the mathematical side of the project and without providing an association of measurements to operators the state determination cannot rightly be called ``empirical''. This measurement-operator assignment is that which we are going to make explicit in what comes. 

\noi
Consider the following measurement-operator assignment in the context of our above model: we associate the following three Bloch vectors to the measurements $a$, $b$ and $c$:
\begin{eqnarray}
{\bf a} &=& {\bf x} = (1,0,0) \label{a}\\
{\bf b} &=& (0,\cos \varphi, -\sin \varphi) \\
{\bf c} &=& {\bf z} = (0,0,1) \label{c}
\end{eqnarray}
and for the sake of simplicity we set the conditional probabilities as follow:
\begin{eqnarray}
p^+_a=p^+_b=p^+_c=:p \label{p}
\end{eqnarray}
The Bloch vector ${\bf s}$ for these special directions and empirical probabilities will then be the following:
\begin{eqnarray}\label{r}
{\bf s}=(p-\frac{1}{2})\left(1, \, \frac{1 + \cos \varphi + \sin \varphi}{1 + \cos \varphi - \sin \varphi}, \, 1 \right)
\end{eqnarray}
But the operator $W_s$ associated to the Bloch vector ${\bf s}$ will not necessarily be a density operator. For example for any 
\begin{eqnarray}
p \in [0.76,1] \quad \mbox{and} \quad \varphi \in [\pi/3, \pi/2) \label{nodens}
\end{eqnarray}
the vector ${\bf s}$ will be longer than 1 and hence $W_s$ will not be positive semidefinite, that is, a density operator. 

Thus, we have provided a noncommutative but not quantum mechanical representation of the above scenario. All the assignments of the table at the beginning of this section hold except the last one: the state of the system is represented by a linear operator but not a density operator. 

This toy-example is, however, special in two senses: (i) the number of measurements is finite and (ii) the number of outcomes is two, that is, the scenario is represented in the two-dimensional Hilbert space which is always a special case. We tackle point (ii) in the next section and point (i) in the one after the next.

\section{Case 2: $k$ measurements with $n$ outcomes}

Let us then see whether a larger set of probabilities can also be given a noncommutative but not quantum mechanical representation. Suppose we perform $k$ measurements on a system such that each measurement can have $n$ outcomes. Suppose we obtain the following empirical conditional probabilities: 
\begin{eqnarray*}
p^j_i := p(A^j_i | a_i)\geqslant 0 \quad  \mbox{with} \quad \sum_i p^j_i =1 \quad   \mbox{for all}  \quad i=1 \dots k; \, j=1 \dots n
\end{eqnarray*}
Just as above we represent each measurement $a_i$ by a self-adjoint operator $O_i$ in the Hilbert space $\H_n$ and the measurement outcomes $\{A^j_i\}$ of $a_i$ by the orthogonal spectral projections $\{P^j_i\}$. The representation is connected to experience by the Born rule: 
\begin{eqnarray*}
 p^j_i := p(A^j_i | a_i) = \mbox{Tr}(W_sP^j_i)
\end{eqnarray*}
where $W_s$ is a linear operator representing the state $s$ of the system. Again, we do not assume that $W_s$ is a density operator; our task is just to see whether it follows that $W_s$ is always a density operator. 

Now, the empirically given probability distributions together with the conventionally chosen sets of minimal orthogonal projections provide constraints on $W_s$ via the Born rule. For a certain number of measurements $W_s$ gets completely fixed. Schematically, 
\begin{eqnarray*}
\{p^j_1\}, \, \{p^j_2\} \, \dots \, \{p^j_k\} \quad \& \quad \{P^j_1\}, \, \{P^j_2\} \, \dots \, \{P^j_k\} \quad \Longrightarrow \quad W_s
\end{eqnarray*}

How many measurements are needed to uniquely fix $W_s$?

$W_s$ gets uniquely fixed if Tr$(W_s A)$ is given for $n^2$ linearly independent operators $A$. Our operators are minimal projections. The first set of minimal orthogonal projections provides $n$ linearly independent equations. Any further linearly independent set of orthogonal projection provides $n-1$ extra equations since in each set the projections sum up to the unity. That is $k$ linearly independent sets of minimal orthogonal projections  provide $k(n-1) +1 $ linearly independent equations which is equal to $n^2$ if $k=n+1$. Thus, performing $k=n+1$ measurements on our system (resulting in $k=n+1$ probability distributions) and representing all the outcomes by orthogonal projections in $\H_n$, the linear operator $W_s$ gets uniquely fixed.

But it will not necessarily be a density operator!

Our question is then: Do $k=n+1$ measurements constrain $W_s$ to be a density operator for \textit{all} linearly independent sets of orthogonal projections representing the outcomes and \textit{all} probability distributions generated from the projections by the Born rule? Again, what we test here is whether a noncommutative representation is necessarily a quantum mechanical representation.

Now, we show that the answer is: \textit{no}. 

As said above, a density operator is a self-adjoint, positive semidefinite operator with trace equal to 1. Self-adjoint operators in $\H_n$ form a vector space $V$ over the field of real numbers. This vector space can also be endowed with an inner product induced by the trace: $(A,B) := \mbox{Tr}(AB)$. The operators with trace equal to 1 form an affin subspace $E$ in $V$ and the positive semidefinite operators form a convex cone $C_+$. (A subset $C$ of a real vector space $V$ that linearly spans $V$ is a \textit{convex cone} if for any $A_1,A_2 \in C$ and $r_1,r_2 \in \RE_+$, $r_1A_1 +r_2A_2 \in C$ and $A, -A\in C\Rightarrow A=0$). The intersection of the two, $\C_+ \cap E$, is a convex set in the affin subspace. The extremal elements of this set are the minimal projections in $\H_n$. Denote this set of minimal projections in $\H_n$ by $\P_n$.

Now, for any cone $C$ in $V$, the \textit{dual cone} $C^*$ is defined as
\begin{eqnarray*}
C^* := \{ A \in V \, | \, \mbox{Tr}(A B) \geqslant 0 \quad \mbox{for all} \, \, B \in C\} 
\end{eqnarray*}
According to Fejér's Trace Theorem the cone of the positive semidefinite operators is self-dual that is $C^*_+= C_+$.

Now, let us return to our example. Consider the $k=n+1$ linearly independent sets of orthogonal projections representing the measurement outcomes in $\H_n$. Let $D$ be the convex cone expanded by these projections in $\P_n$ as extremal elements. Obviously, $D \subset C_+$ and consequently $D^* \supset C^*_+ =C_+$. Pick an element from $(D^* \setminus C_+) \cap E$ and call it $W_s$. Lying outside $C_+$, $W_s$ will \textit{not} be positive semidefinite but, lying in $E$, $W_s$ will be of trace 1. Hence for any set of orthogonal projections it generates a probability distribution by the Born rule. 

Thus, we have found a counter-example (actually, continuously many counter-examples): $k=n+1$ linearly independent sets of orthogonal projections representing measurement outcomes and $k=n+1$ probability distributions such that the latter is generated from the former by the Born rule with an operator $W_s$ which is \textit{not} a density operator (since is not positive semidefinite). Hence, we have provided a noncommutative but not quantum mechanical representation for a situation in which $k=n+1$ measurements with $n$ outcomes are performed on a system. This shows that our previous result is not a consequence of the fact that the Hilbert space is the special $\H_2$. Conditional probabilities of finitely many measurements with finitely many outcomes can always be given a noncommutative but \textit{not} quantum mechanical representation. 

But what is the situation if we are going to the continuum limit? Does our counter-example survive if the cardinality of the set of conditional probabilities to be represented is uncountable? To this we turn in the next section.

\section{Case 3: A continuum set of measurements with $n$ outcomes}

There is a theorem which immediately comes to one's mind when going to the continuum limit, namely Gleason's theorem. 

Suppose we are given a continuum set of probability distributions of measurements with, say, $n$ outcomes. We are to represent this set in an $n$-dimensional Hilbert space $\H_n$. Now, suppose that we assign self-adjoint operators to the measurements such that the spectral projections of the various operators together cover the \textit{full} set $\P_n$ of minimal projections in $\H_n$. In other words, there is no minimal projection in $\P_n$ which does not represent a measurement outcome. In this case we can invoke Gleason's theorem to decide on the question as to whether there exist noncommutative representations which are not the quantum mechanical representation. Gleason's theorem answers this question in the negative. 

Gleason's theorem namely claims that for every state $\phi$ in a Hilbert space with dimension greater than 2 there is a density operator $W$ (and vica versa) such that the Born rule $\phi(P) = \mbox{Tr}(PW)$ holds for \textit{all} projections. In other words, if \textit{all} projections are considered, then the state will uniquely be represented by a density operator. Translating it into our case, the theorem claims that \textit{if} one represents the continuum set of measurement outcomes by the \textit{full} set $\P_n$ of projections of a given Hilbert space, then one has no other choice to account for the whole set of conditional probabilities, than to represent the state by a density operator.

Note, however, that the previous sentence is a conditional: \textit{if} we represent the measurement outcomes by the full set $\P_n$, \textit{then} Gleason's theorem tells us that the only representation is the quantum mechanical. This raises the following question: Are we compelled to represent a continuum set of measurement outcomes necessarily by the full set of minimal projections? Can we not ``compress'' somehow the set of projections representing the measurement outcomes such that (i) the outcome-projection assignment is injective (no two outcomes of different measurements are represented by the same projection), still (ii) the set of projections is only a proper subset of $\P_n$? As we saw in the previous section, in this case we can always represent the state of the system by a linear operator which is not a density operator. Or to put it briefly, can we avoid Gleason's theorem by not making use of all minimal projections of $\P_n$? 

As stressed in Section 2, it is of crucial importance to discern physical measurements from operators mathematically representing them. When we use Gleason's theorem we intuitively assume that \textit{all} projections in a Hilbert space represent a measurement outcome for a real-world physical measurement. The case of spin enforces this intuition since the Bloch sphere representation of spin-half particles nicely pairs the spatial orientations of the Stern-Gerlach apparatus with the projections of $\P_2$. In general, however, we have no \textit{a priori} knowledge of the measurement-operator assignment. Particularly, we cannot assume that a set of measurements \textit{just because it is an uncountable set} has to be represented by the \textit{full} set of projections of a given Hilbert space. \textit{A priori} it is perfectly conceivable that a set of real-world measurements, even if its cardinality is uncountable, can be represented by a proper subset of $\P_n$. 

But now suppose that for a given Hilbert space $\H_n$ \textit{all} the self-adjoint operators on $\H_n$ represent a real-world empirical measurement with $n$ outcomes and \textit{all} states on $\H_n$ represent a real-world preparation of the system to be measured. In other words, take it at face value that the full formalism of an $n$-dimensional quantum mechanics has an empirical meaning. We coin the term \textit{full empirical content of the $n$-dimensional quantum mechanics} for the full (continuum) set of conditional probabilities of measurement outcomes provided by the Born rule, that is by the trace of the product of the different spectral projections and density operators in $\H_n$. Now, our question is this: can the full empirical content of the $n$-dimensional quantum mechanics be represented in $\H_n$ in a noncommutative but \textit{not} quantum mechanical way?

The previous three paragraphs amounts to two different questions, one concerning cardinality the other concerning full empirical content: 
\begin{enumerate}
\item Is a noncommutative representation of a set of empirical probabilities necessarily a quantum mechanical representation if the cardinality of the set is continuum?
\item Is a noncommutative representation of the full empirical content of the $n$-dimensional quantum mechanics necessarily a quantum mechanical representation?                                                                                                                                          \end{enumerate}
In what comes we will show that the answer to the first question is \textit{no} and the answer to the second question is \textit{yes}. 

\noi
We start with the first question. Our task is to represent a continuum set of empirical probabilities in a noncommutative but not quantum mechanical way. The set we pick will be the set of probabilities of spin measurements in all the different spatial directions performed on an electron prepared in \textit{one} given state. This set is obviously a continuum set but not yet the full empirical content of the two-dimensional quantum mechanics since we consider only one state. The continuum set of empirical conditional probabilities is the following:
\begin{eqnarray}
\big\{p^\pm_a := p_s(A^\pm|a); \quad s \, \, \mbox{is fixed}\big\} \label{condprob}
\end{eqnarray}
Here $a$ denotes the spin measurement in direction ${\bf a}$ and $A^\pm$ are the two spin outcomes. Now, in the Bloch sphere representation one associates two unit vectors 
\begin{eqnarray*}
{\bf a} &=&  (1, \vartheta, \varphi) \\
{\bf s} &=&  (1, 0, 0) 
\end{eqnarray*}
to the spin measurement $a$ and state $s$ of the system, respectively, such that the Born rule yields the conditional probabilities (\ref{condprob}):
\begin{table}[h]
\begin{tabular}{ccc} \hline \hline
Operator assignment: && Born rule: \\ \hline \hline 
\begin{tabular}{llr} 
Outcomes: $A^\pm$ & $\longrightarrow$ & $P^\pm_a = \frac{1}{2}(\UN \pm {\bf a}\bm{\s})$  $\quad {\bf a} \in \RE^3, \, |{\bf a}| = 1$ \\
Pure state: $s$ & $\longrightarrow$ & $W_s = \frac{1}{2}(\UN + {\bf s}\bm{\s})$  $\quad {\bf s} \in \RE^3, \, |{\bf s}| = 1$  
\end{tabular}
&&
\begin{tabular}{l} 
$p^\pm_a = \mbox{Tr}(W_sP^\pm_a) $ 
\end{tabular} \\ \hline \hline 
\end{tabular}
\caption{Quantum mechanical representation of the empirical conditional probabilities (\ref{condprob})}
\end{table}

\noindent
As is well-known, the measurement outcomes in the Bloch sphere representation are associated to the \textit{full} set of minimal projections $\P_2$, and hence $W_s$ must be represented by a density operator due to Gleason's theorem. However, the Bloch sphere representation is not the only possible noncommutative representation of (\ref{condprob}). Here is an alternative. 

Consider the following two functions: 
\begin{eqnarray*}
&&f: S^2 \to S^2; \, {\bf a} \mapsto f({\bf a}) \\
&&g: S^2 \to \RE^3; \, {\bf s} \mapsto g({\bf s})
\end{eqnarray*}
and suppose that instead of $\bf a$ and $\bf s$ we associate 
\begin{eqnarray*}
f({\bf a}) &=&  (1, \vartheta', \varphi') \\
g({\bf s}) &=&  (r, 0, 0)
\end{eqnarray*}
to $a$ and $s$, respectively, where
\begin{eqnarray}
\vartheta' &=& \arccos \left(\frac{\cos(\vartheta)}{r} \right) \quad \mbox{for} \, \, \varphi \in [0,2\pi] \label{theta}\\
\varphi' &=& \left\{ \begin{array}{ll} 0 & \mbox{for}\ \vartheta  = 0 \\ \varphi & \mbox{for}\ \vartheta \in (0,\pi) \\ \pi & \mbox{for}\ \vartheta  = \pi \end{array} \right. \label{phi}
\end{eqnarray}
and $r > 1$. Observe that $f$ is injective but not surjective: a spherical cap around the ``North Pole'' and ``South Pole'' is not in the image of $f$. It is easy to check that by these associations we obtain a noncommutative representation for the conditional probabilities (\ref{condprob}):
\begin{table}[h]
\begin{tabular}{ccc} \hline \hline
Operator assignment: && Born rule: \\ \hline \hline 
\begin{tabular}{llr} 
Outcomes: $A^\pm$ & $\longrightarrow$ & $P^\pm_a = \frac{1}{2}(\UN \pm f({\bf a})\bm{\s})$  $\quad f({\bf a}) \in \RE^3, \, |f({\bf a})| = 1$ \\
Pure state: $s$ & $\longrightarrow$ & $W_s = \frac{1}{2}(\UN + g({\bf s})\bm{\s})$  $\quad g({\bf s}) \in \RE^3, \, |g({\bf s})| > 1$  
\end{tabular}
&&
\begin{tabular}{l} 
$p^\pm_a = \mbox{Tr}(W_sP^\pm_a) $ 
\end{tabular} \\ \hline \hline 
\end{tabular}
\caption{A noncommutative but not quantum mechanical representation of the empirical conditional probabilities (\ref{condprob})}
\end{table}

\noindent
The representation is a noncommutative but not a quantum mechanical representation since $W_s$ is not positive semidefinite and hence not a density operator. Note again that we have avoided Gleason's theorem because we did not use the full Bloch sphere to represent measurement outcomes but only a ``belt'' defined by the angles (\ref{theta})-(\ref{phi}). To sum up, even though the set of measurements is uncountable, the noncommutative representation is not necessarily quantum mechanical since the set of projections representing the outcomes is not the full set of projections $\P_2$ of the Hilbert space $\H_2$.

\noi
However, (\ref{condprob}) contains only the conditional probabilities of the spin measurement for \textit{one} state. Can we apply the above technique of ``pecking a hole'' in the surface of the Bloch sphere and ``pushing out'' $s$ such that $W_s$  will not be a density operator in the case when we take into consideration \textit{all states}? In other words, can we provide a noncommutative but not quantum mechanical representation for the full empirical content of the two-dimensional quantum mechanics? This was our second question above.

This is point where the representation of the set of conditional probabilities gets rigid. It will turn out that if one is to represent the conditional probability of \textit{all} measurement outcomes of \textit{all} spin measurement in \textit{all} states, then there is no other noncommutative representation but the quantum mechanical. We prove it by the following lemma. 

\begin{Lemma} \label{lemma 1} Consider the Bloch sphere representation of spin. That is let ${\bf a}$ and ${\bf s}$ two unit vectors  associated to the spin measurement $a$ and state $s$ of the system, respectively, such that the Born rule yields the conditional probabilities: 
\begin{eqnarray} \label{condprob2}
\mbox{Tr}(W_sP^\pm_a) = \mbox{Tr}\left(\frac{1}{2}(\UN + {\bf s}\bm{\s})\frac{1}{2}(\UN \pm {\bf a}\bm{\s})\right) = \frac{1}{2}(1 \pm {\bf sa})
\end{eqnarray}
Then, if there are two functions  
\begin{eqnarray*}
&&f: S^2 \to S^2; \, {\bf a} \mapsto f({\bf a}) \\
&&g: S^2 \to \RE^3; \, {\bf s} \mapsto g({\bf s})
\end{eqnarray*}
such that all the conditional probabilities (\ref{condprob2}) are preserved that is
\begin{eqnarray}
{\bf as} = f({\bf a}) g({\bf s}) \label{scalar_prod_preserve1}
\end{eqnarray}
for all ${\bf a}, {\bf s}\in S^2$, then 
\begin{itemize}
 \item[(i)] $f$ and $g$ are the restrictions of the bijective linear maps
\begin{eqnarray*}
&&\hat f\colon \RE^3 \to \RE^3 \\
&&\hat g\colon \RE^3 \to \RE^3 
\end{eqnarray*}
to $S^2$, respectively;
\item[(ii)] $\hat f$ is the orthogonal transformation;
 \item[(iii)] $\hat g = \hat f$.
\end{itemize}
\end{Lemma}

\noindent
For the proof of Lemma \ref{lemma 1} see the Appendix. 

\noi
Lemma \ref{lemma 1} shows that there is no other transformation of the Bloch vectors which preserve all the empirical conditional probabilities encoded in the inner product but the orthogonal transformation. Consequently, one cannot avoid Gleason's theorem and provide a counter-example of the above type in which the state is represented by a linear but not density operator. 

In the rest of the section we prove that this result holds not only in $\H_2$ but in any $n$-dimensional Hilbert space. We show that one cannot preserve all the empirical conditional probabilities encoded in the inner product of the Hilbert space by other transformation than the unitary transformation. Thus, ``compressing'' the empirical content in a proper subset of $\P_n$ of a given Hilbert space is not a viable route to follow. If all the inner products of minimal projections have an empirical meaning, then the only way to represent them is via quantum mechanics.

\begin{Lemma} \label{lemma 2} Let $\H$ be an $n$-dimensional Hilbert space and let $\P_n$ be the set of minimal projections in $\B(\H)\simeq M_n(\CO)$. If there are two functions  
\begin{eqnarray*}
&&f:\P_n \to \P_n \\
&&g:\P_n \to M_n(\CO)
\end{eqnarray*}
such that 
\begin{eqnarray}
\mbox{Tr}(PQ) = \mbox{Tr}(f(P)g(Q)) \label{scalar_prod_preserve}
\end{eqnarray}
for all $P, Q \in \P_n$, then 
\begin{itemize}
 \item[(i)] $f$ and $g$ are the restrictions of the bijective linear maps
\begin{eqnarray*}
&&\hat f\colon M_n(\CO)\to M_n(\CO) \\
&&\hat g\colon M_n(\CO)\to M_n(\CO)
\end{eqnarray*}
to $\P_n$, respectively;
\item[(ii)] $\hat f$ is unitary with respect to the inner product on $M_n(\CO)$ provided by the trace;
 \item[(iii)] $\hat g = \hat f$.
\end{itemize}
\end{Lemma}

\noindent
For the proof of Lemma \ref{lemma 2} see again the Appendix.\footnote{I thank Péter Vecsernyés for his help in proving both Lemma \ref{lemma 1} and \ref{lemma 2}.}

\section{Discussion}

Is quantum mechanics the only possible way to represent an empirically given set of classical conditional probabilities in a noncommutative way; or is this representation picked out from a broader set of representations by convention? Ultimately, this was the question we posed in this paper. To make this question precise, we specified a set of representations, called noncommutative representations, in which measurement choices and measurement outcomes were represented in the quantum fashion and the Born rule connecting the quantum probabilities to classical conditional probabilities was respected. We asked whether experience can ensure that this representation becomes not just partly but fully quantum mechanical, that is, the state will be represented by a density operator. Our answer was the following: 
\begin{enumerate}
\item In case of finitely many measurements with finitely many outcomes the probability distribution of outcomes can always be given a noncommutative but not quantum mechanical representation. 
\item In case of infinitely many measurements the probability distributions can be given a noncommutative but not quantum mechanical representation only if one can avoid Gleason's theorem by not using all the projections of the Hilbert space in representing measurement outcomes.
\item If the physical situation is so complex that the inner product of any pair of minimal projections is of empirical meaning, then there exists no noncommutative representation which is not quantum mechanical. Thus, the quantum mechanical representation is not conventional.
\end{enumerate}
The relation between point 2 and 3 is very subtle. It shows that simply the cardinality of the set of measurements does not decide on whether the situation can be given a noncommutative but not quantum mechanical representation. By ``compressing'' the projections representing measurement outcomes into a real subset of the full set of minimal projections of the given Hilbert space one can go beyond the quantum mechanical representation. The representation becomes rigid only if the inner product of any pair of minimal projections in a Hilbert space can be given an empirical content. This is the case for spin-half particles where projections can directly be associated to preparations and measurement directions. Whether one can provide a similar empirical account for the inner product of any pair of minimal projections in a Hilbert space of higher dimension, is a question which cannot be decided \textit{a priori}.

\section*{Appendix}

\noi
\textit{Proof of Lemma \ref{lemma 1}.} (i) Let $\{{\bf e}_1,{\bf e}_2, {\bf e}_3\}\subset S^2$ be an orthonormal basis in $\RE^3$. Then due to (\ref{scalar_prod_preserve1}) the sets $\{ f({\bf e}_1), f({\bf e}_2),f({\bf e}_3)\}$ and $\{ g({\bf e}_1), g({\bf e}_2),g({\bf e}_3)\}$ are biorthogonal: 
\begin{eqnarray*}
(f({\bf e}_i),g({\bf e}_j))=\delta_{i,j} \quad i,j=1,2,3
\end{eqnarray*}
Biorthogonal sets with cardinality $d$ in $\RE^d$ form (in general two different) linear bases of $\RE^d$. Hence, if ${\bf a}=\sum_i\alpha_i{\bf e}_i\in S^2$ and $f({\bf a})=\sum_i\alpha_i^f f({\bf e}_i)\in\RE^3$ with $\alpha_i,\alpha_i^f\in\RE$, then 
\begin{equation}
\alpha_i=({\bf a},{\bf e}_i)=(f({\bf a}),g({\bf e}_i))=\sum_j\alpha_j^f(f({\bf e}_j),g({\bf e}_i))=\alpha_i^f,\quad i=1,2,3
\end{equation}
Hence, $f(\sum_i\alpha_i{\bf e}_i)=\sum_i\alpha_if({\bf e}_i)$, that is $f$ is the restriction of the bijective linear map $\hat f$ characterized by the image linear basis $\{ f({\bf e}_1),f({\bf e}_2,f({\bf e}_3)\}$ of the orthonormal basis $\{ {\bf e}_1,{\bf e}_2,{\bf e}_3\}$. A similar argument shows that $g$ is the restriction of the bijective linear map $\hat g$ to $S^2$.

\noi
(ii) Using polarization identity 
\begin{eqnarray*}
({\bf a},{\bf b})=\frac{1}{4}\big[({\bf a}+{\bf b},{\bf a}+{\bf b})-({\bf a}+{\bf b},{\bf a}+{\bf b})\big],\quad {\bf a},{\bf b}\in \RE^3
\end{eqnarray*}
it is enough to show that 
\begin{eqnarray*}
({\bf a},{\bf a})=(\hat f({\bf a}),\hat f({\bf a})),\quad {\bf a} \in \RE^3
\end{eqnarray*}
which, however, holds since
\begin{eqnarray*}
1=({\bf a},{\bf a})=(f({\bf a}),f({\bf a}))=(\hat f({\bf a}),\hat f({\bf a})),\quad {\bf a} \in S^2
\end{eqnarray*}
and $\hat f$ is linear.

\noi
(iii) Using (\ref{scalar_prod_preserve1}) and the orthogonality of $\hat f$ one has \[ ({\bf a},{\bf b})=(\hat f({\bf a}),\hat g({\bf b}))=({\bf a},\hat f^{-1}(\hat g({\bf b}))),\ {\bf a},{\bf b}\in \RE^3. \] Hence, $\hat g=\hat f$ due to the uniqueness of the inverse map.\qed

\noii
\textit{Proof of Lemma \ref{lemma 2}.} (i) Since the trace is a faithful positive linear functional on $M_n(\CO)$,
\[ (A,B):=\mbox{Tr}(A^*B),\quad A,B\in M_n(\CO)
\]
defines an inner product on the $n^2$-dimensional complex linear space $M_n(\CO)$. The real linear combinations of the projections in $\P_n$ span the real vector space of self-adjoint elements in $M_n(\CO)$, and the complex linear combinations span the complex vector space $M_n(\CO)$. Let $\{P_i, i=1,\dots, n^2\}\subset\P_n$ be a linear basis in $M_n(\CO)$. Then the inner product matrix $g\in M_{n^2}(\CO)$ given by matrix elements $g_{ij}:=(P_i,P_j)\geq 0$ is an invertible matrix. Since
$\mbox{Tr}(f(P_i)g(P_j))=g_{ij}$ due to (\ref{scalar_prod_preserve})  $\{f(P_i), i=1,\dots, n^2\}\subset\P_n$ and $\{g(P_i), i=1,\dots, n^2\}\subset M_n(\CO)$ are linear bases in $M_n(\CO)$ due to invertibility of $g$. Defining the bijective linear maps $\hat f,\hat g\colon M_n(\CO)\to M_n(\CO)$ by the linear extension of these image bases for $P=\sum_i\alpha_iP_i\in\P_n$ one has 
\begin{eqnarray}
(f(P),g(P_j))=(P,P_j)=(\sum_i\alpha_iP_i,P_j)=\sum_i\alpha_i(P_i,P_j) =\sum_i\alpha_i(f(P_i),g(P_j))\nonumber\\ 
=(\sum_i\alpha_if(P_i),g(P_j)) =:(\hat f(\sum_i\alpha_iP_i),g(P_j))=(\hat f(P),g(P_j)),\quad j=1,\dots, n^2.
\nonumber
\end{eqnarray}
Hence, $f$ is the restriction of the bijective linear map $\hat f$ to $\P_n$, indeed. A similar argument shows that $g$ is the restriction of the bijective linear map $\hat g$ to $\P_n$.

\noi
(ii) Using polarization identity 
\begin{eqnarray*}
(A,B)=\frac{1}{4}\big[(A+B,A+B)-(A-B,A-B)\big],\quad A,B\in M_n(\CO)
\end{eqnarray*}
it is enough to show unitarity on `diagonal' inner products: 
\begin{eqnarray*}
(A,A)=(\hat f(A),\hat f(A)),\quad A\in M_n(\CO)
\end{eqnarray*}
Since $f(\P_n)\subset \P_n$ by assumption, using the normalization $\mbox{Tr}(P)=1, P\in\P_n$ of the trace it follows that
\begin{eqnarray*}
(P,P)=1= (f(P),f(P)) = (\hat f(P), \hat f(P)),\ P\in\P_n
\end{eqnarray*}
i.e. $\hat f$ is unitary on diagonals from $\P_n$. Using a spectral decomposition of self-adjoint elements by orthogonal minimal
projections one concludes that $\hat f$ maps the real vector space of self-adjoint elements in $M_n(\CO)$ into itself, moreover, it is unitary on diagonals from the space of self-adjoint elements. Since $A\in M_n(\CO)$ can be written uniquely as a sum of self-adjoint elements: $A=R+iI$ with $R:=(A+A^*)/2$ and $I:=(A-A^*)/2i$ it follows that
\begin{eqnarray*}
(A,A)&=&(R+iI,R+iI)=(R,R)+(I,I)=(\hat f(R),\hat f(R))+(\hat f(I),\hat f(I))\\
&=&(\hat f(R)+i\hat f(I),\hat f(R)+i\hat f(I)) = (\hat f(A),\hat f(A)),
\end{eqnarray*}
that is $\hat f$ is unitary on diagonals from $M_n(\CO)$, which provides unitarity of $\hat f$.

\noi
(iii) Using (\ref{scalar_prod_preserve}) and unitarity of $\hat f$ one has 
\[ (A,B)=(\hat f(A),\hat g(B))=(A,\hat f^{-1}(\hat g(B))),\ A,B\in M_n(\CO). \]
Hence, $\hat g=\hat f$ due to the uniqueness of the inverse map.\qed
\vspace{0.2in}

\noindent
{\bf Acknowledgements.} I thank Zalán Gyenis, Sam Fletcher and especially Péter Vecsernyés for valuable discussions. This work has been supported by the Hungarian Scientific Research Fund, OTKA K-115593.

\section*{References}
\small
\begin{list} 
{ }{\setlength{\itemindent}{-15pt}
\setlength{\leftmargin}{15pt}}

\item G. Bana and T. Durt, ''Proof of Kolmogorovian Censorship,'' \textit{Found. Phys.}, \textbf{27}, 1355–1373. (1997). 

\item G. Chiribella, G. M. D'Ariano and P. Perinotti, \textit{Quantum Theory from First Principles}, (Cambridge: Cambridge University Press, 2017).

\item L. E. Szabó, ''Is quantum mechanics compatible with a deterministic universe? Two interpretations of quantum probabilities,'' \textit{Foundations of Physics Letters}, \textbf{8}, 417-436 (1995).

\item L. E. Szabó, ''Critical reflections on quantum probability theory,'' in M. Rédei and M. Stöltzner (eds.) \textit{John von Neumann and the Foundations of Quantum Physics}. (Institute Vienna Circle Yearbook. Dordrecht: Kluwer Academic Publishers, 201–219, 2001)

\item L. E. Szabó, ``The Einstein-Podolsky-Rosen argument and the Bell inequalities,'' \textit{Internet Encyclopedia of Philosophy}, URL= http://www.iep.utm.edu/epr/ (2008).

\item G. M. D’Ariano, L. Maccone and M.G. A. Paris, ``Quorum of observables for universal quantum estimation,''  J. Phys. A, \textbf{34}, 93-103, 2001.

\item S. Gudder, \textit{Quantum Probability}, (Academic Press, San Diego, 1988)

\item L. Hardy, ``Quantum Theory From Five Reasonable Axioms,'' arxiv.org/abs/quant-ph/0101012, 2008.

\item J. L. Park and W. Band, ``The empirical determination of quantum states,'' \textit{Found. Phys.}, \textbf{1}, 133-144 (1970).

\item J. L. Park and W. Band, ``A general theory of empirical state determination in quantum mechanics, Part I and Part II,'' \textit{Found. Phys.}, \textbf{1}, 211-226 and 339-357  (1971).

%\item J. L. Park and W. Band, ``Preparation and Measurement in Quantum Physics,'' \textit{Found. Phys.}, \textbf{22}, 657-668 (1992).

\item M. Rédei, ''Kolmogorovian Censorship Hypothesis for general quantum probability theories,'' \textit{Manuscrito - Revista Internacional de Filosofia}, \textbf{33}, 365–380 (2010).

\item M. Rédei and S. Summers ''Quantum probability theory,'' \textit{Stud. Hist. Phil. Mod. Phys.}, \textbf{38}, 390-417 (2007).

\end{list}
\end{document}